\documentclass[apj]{emulateapj}
\usepackage{natbib,graphicx,amsmath,amsthm,ulem,color}

\def\Kepler{\textit{Kepler~}}

\shorttitle{Model-Independent Stellar and Planetary Masses}
\shortauthors{Montet and Johnson}

\begin{document}

\title{Model-Independent Stellar and Planetary Masses From Multi-Transiting Exoplanetary Systems}
\author{Benjamin T. Montet\altaffilmark{1} and John Asher Johnson\altaffilmark{2,3}}
\email{btm@astro.caltech.edu}
\altaffiltext{1}{Cahill Center for Astronomy and Astrophysics, California Institute of Technology, 1200 E. California Blvd., MC 249-17, Pasadena, CA 91125, USA}
\altaffiltext{2}{Division of Geological and Planetary Sciences, California Institute of Technology, 1200 E. California Blvd., MC 170-25, Pasadena, CA 91125, USA}
\altaffiltext{3}{NASA Exoplanet Science Institute (NExScI), CIT Mail Code 100-22, 770 South Wilson Avenue, Pasadena, CA 91125, USA}

\begin{abstract}

  Precise exoplanet characterization requires precise classification of exoplanet host stars. The masses of host stars are commonly estimated by comparing their spectra to those predicted by stellar evolution models. However, spectroscopically determined properties are difficult to measure accurately for stars that are substantially different from the Sun, such as M-dwarfs and evolved stars. Here, we propose a new method to dynamically measure the masses of transiting planets near mean-motion resonances and their host stars by combining observations of transit timing variations with radial velocity measurements. We derive expressions to analytically determine the mass of each member of the system and demonstrate the technique on the Kepler-18 system. We compare these analytic results to numerical simulations and find the two are consistent. We identify eight systems for which our technique could be applied if follow-up radial velocity measurements are collected. We conclude this analysis would be optimal for systems discovered by next generation missions similar to TESS or PLATO, which will target bright stars that are amenable to efficient RV follow-up.

\end{abstract}

\keywords{Methods: analytical; Planets and satellites: fundamental parameters; Stars: fundamental parameters}

\section{Introduction}
\label{S:Intro}
With modern radial velocity techniques and the phenomenal success of space-based transit surveys, exoplanetary science has moved from a ``stamp-collecting'' era of finding individual systems to an era where hundreds of planetary systems are discovered simultaneously \citep{Borucki11}. Despite these successes, accurate characterization of planets still presents a challenge. In general, uncertainties in the radii and masses of planets are dominated by uncertainties in the radii and masses of their host stars \citep[e.g.][]{KOI961}. Difficulties in characterizing the physical properties of planets are particularly acute for systems discovered by the \Kepler space telescope. For many systems, the ratio between the radius of the planet and the radius of its host star is known to within 1 part in 1000 \citep{Batalha12}. Yet the stellar radii are often not known to better than ten percent, meaning much of the precision of \Kepler is lost when estimating planetary properties \citep{Lissauer12, KOI254}.

In general, measuring the masses of exoplanet host stars is a model-dependent procedure. For nearby stars with trigonometric parallaxes, one compares the luminosity, effective temperature, and metallicity of a star to stellar evolution model grids \citep{ValFis05, JohnsonMW12}. For stars without measured parallaxes, the stellar density can be measured from the transit light curve and used in place of the luminosity. However, this relies on either the assumption that the planet's orbit is circular---a poor assumption for periods larger than 10 days---or an RV orbital solution \citep{Sozzetti07, DawsonJJ12}. The atmospheres and interior structures of stars are also poorly understood for stars that differ substantially from the Sun, complicating their analyses further. Thus, model-independent methods of measuring stellar masses are extremely valuable. 

\citet{Agol05} suggest that in a system with transiting planets, a precise measurement of the transit duration, which depends on stellar density, coupled with radial velocity information and precise measurements of the scatter in transit times can provide a unique measurement of the stellar mass. Unfortunately, this strategy requires precise knowledge of the inclination of the system, which from a transit light curve is degenerate with limb-darkening coefficients \citep{Jha00}, especially for low signal-to-noise transit detections.

The method described by Agol et al.\ also breaks down for resonant systems, as it assumes the relative positions of the planets change from transit to transit. Moreover, outside of resonance, transit timing effects are small for all but the largest planets, so this method is not ideal for studying rocky planets. This strategy is successful when the perturbing object is massive, as is the case in circumbinary planets \citep{Doyle11, Welsh12} but is less promising for studying solar-type systems. It has also been suggested that in a system containing a transiting planet and an exomoon detected through transit timing and duration variations, the stellar mass and radius can be determined directly through dynamical effects \citep{Kipping10}. While this technique holds future promise, the exomoons required to test this procedure have not yet been detected.

Recently, transit timing variations caused by mutual gravitational interactions of bodies in multiple-planet systems have been detected \citep{Holman10, Ford12TTV}. These deviations from a linear transit ephemeris allow for an estimate of the ratio of the mass of the perturbing planet to the mass of its star. In cases where multiple planets transit, the ratio of the masses of each planet to the mass of the host star can be estimated \citep{Fabrycky12, Steffen12a}.

In this paper, we propose a method to directly measure stellar and planetary masses for multi-transiting systems by combining an analysis of the transit timing signal caused by planet-planet interactions with Doppler radial velocity measurements. Unlike the technique developed by \citet{Agol05}, our method requires the observed transiting planets to lie near a mean-motion resonance, where transit timing effects are strongest. In \S2, we explain how transit timing variations can be combined with radial velocity information to estimate stellar and planetary masses. In \S3, we apply our process to the well-studied Kepler-18 planetary system, and compare the result to both numerical integrations of the system and published stellar evolution models. We find the scheme is viable, but at present the radial velocity data is insufficient to place meaningful constraints on stellar parameters. In \S4, we discuss uncertainties and limitations to our method, as well as its applications to systems discovered by \Kepler and its eventual successors.

\section{Unique Masses and Errors}

\subsection{Mass Determination from TTVs}

Consider a system of two coplanar planets orbiting near (but not exactly at) a first-order mean motion resonance. The planets have periods $P$ and $P^\prime$ (here and throughout, the unprimed quantity refers to the inner planet and the primed quantity to the outer planet) and orbit their star such that the inner planet completes approximately $j$ orbits in the time the outer planet completes $j-1$. A nearly edge-on observer will detect both planets transiting their host star. Because of the near-commensurability of their periods, the inner planet will pass its companion at nearly the same location each orbit, driving small gravitational interactions which add coherently, inducing a small “forced eccentricity” on each object. The two planets will therefore not transit their star in an exactly periodic fashion. Instead, a small, sinusoidal departure from periodicity, termed a transit timing variation (TTV), will be observed \citep[e.g.][]{Nesvorny12}. TTVs have been used to detect the presence of nontransiting planets \citep{Ballard11, Dawson12} and to fully characterize systems when multiple planets transit \citep{Holman10, Lissauer11}. The period of the TTV signal is related to the periods of the planets such that
\begin{equation}
P_{TTV} = \frac{1}{|j / P^\prime - (j-1) / P|}.
\end{equation}
In most cases, the superb photometry provided by the \Kepler mission allows this quantity to be precisely estimated. 

An analytic form for the amplitude of the TTV signal is derived by \citet[hereafter L12]{Lithwick12}. The amplitude of the signal depends strongly on the free eccentricity of the system. Here, free eccentricity refers to the component of the eccentricity caused by the initial dynamical conditions of the system, not the component driven by resonant interactions. Without observing a secondary transit, for small planets the free eccentricity is difficult to constrain precisely via photometry. However, many TTV signals have been detected in systems in which the planets have orbital periods of days to weeks. 

In this case, the estimated ages of the planets are larger than the expected tidal circularization timescale at their present locations, so their orbits can be expected to have negligible free eccentricity. This can be verified by analyzing the phase of the TTV signal. If the zero points of the TTV signal occur when the longitude of conjunction is parallel to the line of sight, L12 suggest the free eccentricity can be neglected. In this case, the amplitudes of the TTV signals, $V$ and $V^\prime$, are
\begin{align}
\label{Veqn2}
V &= \frac{m^\prime}{M} \bigg| \frac{f}{\Delta} \bigg| \frac{P}{\pi j^{2/3} (j-1)^{1/3}} \\
V^\prime &= \frac{m}{M} \bigg| \frac{g}{\Delta} \bigg| \frac{P^\prime}{\pi j} 
\label{Veqn}
\end{align}
where $m$ and $M$ are the planet and stellar mass, $\Delta$ is the fractional distance from commensurability, typically of order 0.01, and $f$ and $g$ the appropriate coefficient of the disturbing function, which characterizes the interactions between the planets. These sums of Laplace coefficients can be calculated by using the information found in Appendix B of \citet{MDSSD}. Additionally, the values of $f$ and $g$ for common resonances are listed in L12. To first order, these coefficients are of order unity and depend only weakly on $\Delta$. For systems with TTVs, Equations \ref{Veqn2} and \ref{Veqn} enable a unique determination of the planet-star mass ratio, but normally one must rely on stellar models to separate stellar and planetary properties. However, if radial velocity measurements are available, the amplitude of the Doppler signal can be used in conjunction with the TTV information to estimate the masses of the planets and the star.

\subsection{Including Radial Velocities}
The semiamplitude of a radial velocity Doppler signal is
\begin{equation}
K = \bigg( \frac{2\pi G}{P} \bigg) ^{1/3} \frac{m}{(M+m) ^{2/3}} \frac{\sin i}{\sqrt{1-e^2}},
\end{equation}
with $i$ and $e$ the inclination and eccentricity, respectively \citep{Paddock13}. Despite the lack of free eccentricity, we may expect a small forced eccentricity as a result of 3-body interactions. The magnitude of this forced eccentricity is $\lesssim 0.05$, so neglecting it will induce an error of $\lesssim 0.1$\%  in our semiamplitude calculation. An error of the same magnitude but in the opposite direction is induced by assuming $ i = 90 ^\circ$, since a strong constraint on the inclination is provided by our requirement of a transit.

Thus, neglecting eccentricity and assuming an edge-on orbit, in the limit where $m \ll M$ the radial velocity semiamplitude can be approximated as
\begin{equation}
K = \bigg (\frac{2\pi G}{P} \bigg) ^{1/3} \frac{m}{M ^{2/3}},
\label{Keqn}
\end{equation}
and the radial velocity can be modeled as
\begin{equation}
v(t) = -K \sin\bigg(\frac{2\pi(t - t_c)}{P}\bigg),
\label{RVeqn}
\end{equation}
with $t_c$ the time of transit center. Again, a degeneracy exists between the planet and stellar mass, so stellar models must be invoked. However, the degeneracy is different from the one recovered from transit timing variations, so these two expressions taken together can be used to solve for the planet and stellar masses individually. This allows for two independent measurements of the mass of the star and one unique measurement of the mass of each planet. The mass of the star is

\begin{align} 
M &= \bigg[ \frac{P P^{\prime 3} g^3}{2 \pi^4 G \Delta^3 j^3} \bigg] \frac{K^3}{V^{\prime 3}} \\
  &= \bigg[ \frac{P^\prime P^3 f^3}{2 \pi^4 G \Delta^3 j^2 (j-1)} \bigg] \frac{K^{\prime 3}}{V^3},
\end{align}
and the mass of each planet is

\begin{align}
m &= \frac{P^3}{2 \pi G} \bigg(\frac{P^\prime g}{\Delta \pi j}\bigg)^2 \frac{K^3}{V^{\prime 2}} \\
m^\prime &= \frac{P^{\prime 3}}{2 \pi G} \bigg(\frac{P f}{\Delta \pi j^{2/3} (j-1)^{1/3}}\bigg)^2 \frac{K^{\prime 3}}{V^2}.
\end{align}

Simply put, for a given system of two coplanar planets, the constants $M^{1/3}V^{\prime}K^{-1}$ and $M^{1/3}V K^{\prime -1}$ depend only on the system architecture. Thus, by precisely measuring the RV semiamplitude and magnitude of the TTV signal, the stellar mass can be directly estimated. Because of \textit{Kepler}'s exceptional photometry, the periods of each planet and terms derived from these (such as $\Delta$) are well known. Thus we expect the errors in the mass estimates to be dominated by the errors in $V$ and $K$, and neglect the errors caused by other terms. In this case,

\begin{align}
\bigg(\frac{\delta M}{M}\bigg)^2 &=  \bigg(3 \frac{ \delta K }{K}\bigg) ^2 +  \bigg(3 \frac{ \delta V^\prime}{V^\prime} \bigg) ^2 \nonumber \\
 &=  \bigg(3\frac{ \delta K^\prime }{K^\prime}\bigg) ^2 +  \bigg(3 \frac{\delta V}{V} \bigg) ^2 \\
\bigg(\frac{\delta m}{m}\bigg)^2 &=  \bigg(3\frac{ \delta K }{K}\bigg) ^2 +  \bigg(2 \frac{\delta V^\prime}{V^\prime} \bigg) ^2 \\
\bigg(\frac{\delta m^\prime}{m^\prime}\bigg)^2 &=  \bigg(3\frac{ \delta K^\prime }{K^\prime}\bigg) ^2 +  \bigg(2 \frac{\delta V}{V} \bigg) ^2,
\end{align}
where we expect the covariant terms to be zero since $K$ and $V$ are independently measured quantities. Here, the fractional uncertainties depend quite sensitively on the ability to measure $K$ and $V$. Typically for systems of multi-transiting planets, only one of these quantities is well measured. Therefore, we would expect to only weakly constrain the stellar masses at present; with more observations the constraints will tighten considerably. In the limit where $m \gg m^\prime$, $K$ and $V^{\prime}$ are much larger than their counterparts and can be more easily constrained. Thus when one planet is substantially more massive than its companion, one stellar mass measurement will be considerably more precise than the other and the mass of the more massive planet will be better constrained than the less massive planet. In the case where $m \approx m^\prime$, both measurements are expected to have similar uncertainties. 

\section{Example}

Kepler-18 (KOI 137, KIC 8644288) is a planetary system containing three nearly coplanar planets with 3.5, 7.6, and 14.9 day periods orbiting a $0.97 M_\odot$ star \citep[henceforth C11]{Cochran11}. These planets (137.03, 137.01, and 137.02, or Kepler-18 b, c, and d, respectively) were confirmed by a combination of transit timing and radial velocity measurements. The star has been observed using the \Kepler short cadence mode nearly continuously for two years, allowing for precise measurements of transit times over dozens of transits. Moreover, 18 radial velocity measurements of this $K_p = 13.5$ star, where $K_p$ is the apparent magnitude in the \Kepler bandpass, have been collected over the past three years by the California Planet Search team with the Keck 1 High Resolution Echelle Spectrometer (HIRES). Thus, enough data exist to attempt to determine the mass of each member of this system dynamically.

We first fit a limb-darkened light curve to a series of phase-folded transits to estimate the observable transit parameters, such as the impact parameter and limb-darkening coefficients, following the OCCULTQUAD routine developed by \citet{Mandel02}. Because of the high signal to noise ratio of these observations and the one-minute integration times, transit parameters can be easily measured from individual transits: we find no significant difference in these parameters or their uncertainties when fitting one individual transit instead of fitting a phase-folded transit. 

Once the shape of the light curve is modeled, we fit a curve of this shape to each individual transit, allowing only the time of transit center to vary. Each individual transit light curve consists of over 200 in-transit data points, allowing for measurements of the transit center time to sub-minute precision. We remove from our dataset transits that occur simultaneously with the transit of another planet. As expected, the transits follow a sinusoidal deviation from a linear ephemeris; these deviations, shown in Table \ref{TTTable}, appear to be anti-correlated between the two planets.

For our method to provide meaningful mass estimates, the primordial (free) eccentricity of the system must be damped on a timescale shorter than the age of the system. As explained in L12, if the zero points of the transit timing variations occur at the times at which the longitude of conjunction of the planets is equal to 0 or 180 degrees, then the system is likely to have negligible free eccentricity. We check the phase of these transit timing variations by fitting each TTV curve independently to a sinusoid. We determine parameters of this sinusoid and their uncertainties by minimizing the $\chi^2$ statistic through a Levenberg-Marquardt algorithm. Additionally, we allow for a vertical offset to the sine function in the fit (indicative of a miscalculated time of transit center $t_c$), and also a linear trend (indicative of a miscalculated orbital period). The results of this minimization can be found in Table \ref{TTVdata}. Both planets are consistent with having zero free eccentricity and anticorrelated TTV signals. From these parameters, we measure a fractional distance from commensurability $\Delta = -2.776 \times 10^{-2}$ and find the coefficients of the disturbing function to be $f = -1.251$ and $g = 0.5308$. The amplitude of the TTV signals for Kepler-18 c and d can be measured to within 3.3 and 6.8 percent, respectively. 

\begin{deluxetable*}{cccccc}
	\tabletypesize{\footnotesize}
	\tablecolumns{6}
        \tablecaption{TTV Fitting Results for Kepler-18}
        \tablehead{ \colhead{Planet} & \colhead{Phase (deg)} & \colhead{Amplitude (min)} & \colhead{Period (day)} & \colhead{$T_c$ (BJD - 2454900.0)} & \colhead{TTV Period (day)}}
        \startdata
        18-c & $184.5 \pm 4.1$ & $5.54 \pm 0.18$ & 7.6415716(5) & 68.4071(2) & $265.1 \pm 2.5$ \\
        18-d & $3.2 \pm 8.8$ & $4.46 \pm 0.30$ & 14.858941(1) & 61.1531(1) & $265.9 \pm 5.3$
        \enddata
        \label{TTVdata}
\end{deluxetable*}
These results can then be combined with radial velocity measurements in order to uniquely constrain the stellar and planetary masses. C11 used 14 radial velocity measurements to confirm this system; we analyze these data plus four additional observations collected between 1 July 2012 and 1 August 2012, all of which are provided in Table \ref{RVels}.

\begin{deluxetable}{ccc}
\tablewidth{0pt}
\tabletypesize{\footnotesize}
\tablecaption{Keck HIRES Relative Radial Velocity Measurements of Kepler-18.}

\tablehead{\colhead{BJD-2440000} & \colhead{RV}           & \colhead{$\sigma$} \\
                        & \colhead {m\,s$^{-1}$} & \colhead {m\,s$^{-1}$} }
\startdata
 15076.009  & 7.750   & 2.539\\
 15076.927  &  6.950  & 2.487\\
 15081.024  &  8.617  & 4.214\\
 15082.007  & -1.007  & 2.381\\
 15084.984  & -7.320  & 2.977\\
 15318.066  &  3.388  & 2.625\\
 15322.029  & -10.093 & 2.303\\
 15373.004  &  12.189 & 2.150\\
 15403.019  &  24.983 & 2.915\\
 15405.909  & -11.692 & 2.350\\
 15406.881  &  0.340  & 2.195\\
 15413.011  & -10.788 & 2.498\\
 15432.970  &  0.205  & 2.233\\
 15436.782  & -6.675  & 2.256\\
 16109.905  & -14.919 & 2.261\\
 16111.845  &  4.123  & 2.642\\
 16115.973  & -0.326  & 2.434\\
 16140.839  & -8.153  & 2.730

\enddata
\label{RVels}
\end{deluxetable}

The large uncertainties in each individual observation, coupled with the small number of observations relative to the number of observed transits, suggest our mass uncertainties will be dominated by uncertainties in the radial velocity semiamplitude. In fact, many different solutions fit the RV data satisfactorily. As an example, C11 fit a larger RV semiamplitude for planet d than c, despite the fact that they find planet c to be both more massive and nearer the star than planet d. The analysis is complicated by the existence of the much smaller planet b orbiting inside the other two planets. In this case, we invoke one additional piece of information. Equations \ref{Veqn2} and \ref{Veqn} can be combined to solve for the mass ratio of the resonant planets, 
\begin{equation}
\frac{m^\prime}{m} = \frac{P}{P^\prime} \frac{f}{g} \frac{V^\prime}{V} \bigg(\frac{j}{j-1}\bigg)^{1/3},
\label{Mratio}
\end{equation}
which in this case implies $m^\prime/m = 1.22 \pm 0.09$, where $m^\prime$ refers to planet c and $m$ to planet d. This can be applied as an additional constraint in the radial velocity fit. In the case where $\sigma_{RV} \ll \sigma_{TTV}$, an equivalent mass ratio constraint, derived from the radial velocity semiamplitude ratio, can be applied to the TTV fit. 

With this additional constraint, we model the RVs as the sum of three sinusoids of the form of Equation \ref{RVeqn}, with three free parameters: the semiamplitude of one of the resonant planets (c or d; here, we fit c), the semiamplitude of the innermost planet b, and an offset term, $\gamma$. We find the best fitting parameters to be $K_c = 6.89 \pm 1.40$ m s$^{-1}$, $K_b = 4.18 \pm 2.14$ m s$^{-1}$, and $\gamma = 1.30 \pm 1.45$ m s$^{-1}$. From the mass ratio above, this implies a semiamplitude for planet d of $K_d = 4.52 \pm 0.97$ m s$^{-1}$. We now have enough information to estimate the stellar and planetary masses; these results are shown in Table \ref{ExampRes}. 

\begin{deluxetable*}{ccccc}
	\tabletypesize{\footnotesize}
	\tablecolumns{5}
        \tablewidth{0pt}
        \tablecaption{Mass Estimates for Kepler-18 system}
        \tablehead{ \colhead{Object} & \colhead{C11} & \colhead{L12} & \colhead{Analytic Result\tablenotemark{a}} & \colhead{Dynamical Estimate\tablenotemark{b}}}
        \startdata
        Star ($M_\odot$) & $0.972 \pm 0.042$ & Assumed C11 &  $0.83 \pm 0.51$ & $0.92 ^{+ 0.61} _{-0.40}$ \\
        Planet c ($M_\oplus$) & $17.3 \pm 1.8$ & $20.2 \pm 1.9$ & $18.6 \pm 11.6 $ & $14.8 ^{+9.4} _{-6.0}$ \\
        Planet d ($M_\oplus$) & $16.4 \pm 1.4 $ & $17.4\pm 1.2$ & $15.4 \pm 9.5 $ & $15.4 ^{+11.0} _{-7.0}$ 
        \enddata 
        
        \tablenotetext{a}{Result derived by applying Equations 11-13}
        \tablenotetext{b}{Result determined from numerical integrations}
   
        \label{ExampRes}
\end{deluxetable*}

When both the RV and TTV amplitudes are measured without invoking the extra constraint of Equation \ref{Mratio}, two independent measurements of the stellar mass can be calculated, one through $K$ and $V^\prime$, and one through $K^\prime$ and $V$. However, since our value for $K^\prime$ is found by assuming a value for $K$, we only calculate one independent measure of the stellar mass. We find a stellar mass of $0.83 \pm 0.51 M_\odot$, consistent with that found by C11. We find the masses of planets c and d to be $18.6 \pm 11.6$ and $15.4 \pm 9.5 M_\oplus$, respectively.

It is somewhat disappointing that the uncertainties in the stellar mass are so large in this example, but this should be considered a shortcoming in the available data, not in the potential of our technique. Because most Kepler Objects of Interest (KOIs) are considerably fainter than typical stars probed by radial velocity surveys, follow-up radial velocity measurements are often carried out only to a level necessary to confirm the planetary nature of a transiting system, if any observations are collected at all. Thus for most systems that exhibit transit timing variations, radial velocity measurements alone are rarely precise to within even 20\%. Better constraints on $K$ are regularly achieved for stars targeted in radial velocity surveys, and with more follow-up observations these mass estimates will be greatly improved. This is discussed more fully in \S\ref{S:SD}.

We can confirm the validity of our method by comparing our analytic result to results obtained through numerical integrations of this system. To accomplish this task, we make use of the Systemic Console developed by \citet{Meschiari09}. This program is designed to simultaneously fit Doppler and transit timing measurements. The Console contains several built-in integrators, including an eighth-order Runge-Kutta scheme employed in this work. The Console is not designed to enable the user to solve for the stellar mass as a free parameter. We circumvent this problem by first assuming a stellar mass. We fix the period and mean anomaly at BJD = 2455128.0 so that they are consistent with values found in Table 7 of C11; we then allow the planet masses, inclinations, and eccentricities to vary and minimize the $\chi^2$ of the system. Once $\chi^2$ is calculated, we vary the stellar mass slightly and repeat this procedure. With this technique, we can map the likelihood space in both $M$ and $m$. As shown in Table \ref{ExampRes}, both the best fitting parameters and their uncertainties are consistent with the analytic result, suggesting that our method is viable and that dynamical techniques can be used in conjunction with our analytic result to further constrain the stellar and planetary parameters. 

\begin{figure}[htbp]
\centerline{\includegraphics[width=0.5\textwidth]{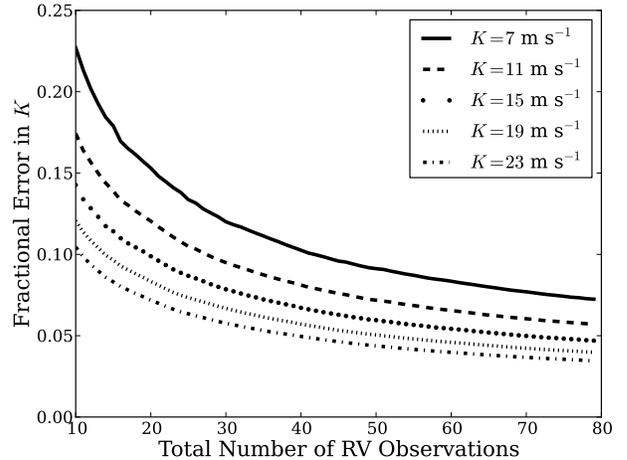}}
\caption{ Derived fractional errors in the Doppler semiamplitude measurement as a function of number of radial velocity observations taken, for various values of $K$, the Doppler semiamplitude. For all observations, the same statistical uncertainties and RV jitter levels are assumed. The jitter level is 3 m s$^{-1}$, a reasonable estimate for all but the youngest dwarf stars. From top to bottom, these curves represent semiamplitudes of $[7, 11, 15, 19, 23]$ m s$^{-1}$. For a system like Kepler-18, where $K \approx 7$ m s$^{-1}$, many more measurements would be required to constrain $K$ to five percent (and thus the stellar mass to 15 percent). However, for a system with either larger planets or a smaller star, this level of precision could be reached with fewer observations.
  }

\label{RVerrors}
\end{figure}

In all cases, our uncertainties are dominated by our $20 \%$ errors in the radial velocity semiamplitudes. The uncertainty in the radial velocity semiamplitude will decrease considerably with more radial velocity observations. We prove this claim by simulating observations placed randomly between the months of June and October, when the \Kepler field is visible at night. We first find the true radial velocity of the system at that time, assuming $K_c = 7.0$ m s$^{-1}$. An statistical uncertainty $\sigma$ is randomly drawn from the observed errors in previous HIRES measurements, and a Gaussian random number is drawn from a distribution $\mathcal{N}(0,\sigma)$. The radial velocity measurement is shifted by an amount equal to this random number, and the statistical uncertainty is recorded as $\sigma$. Finally, to simulate the effects of radial velocity ``jitter'' caused by stellar pulsations, a random number is drawn from $\mathcal{N}(0,3 $ m s$^{-1})$; this value is also added to the radial velocity measurement. The observations are fitted to a combination of sinusoids as described above, and the stellar and planetary masses are estimated. The fractional error in the semiamplitudes for the largest planet as a function of the number of observations is shown in Fig. \ref{RVerrors} (solid line).

We find that, with 30 more radial velocity observations, the uncertainty in our calculation drops by nearly a factor of two, from the current 61 percent to 33 percent. To provide substantially better than 33 percent uncertainties without obtaining 50 radial velocity measurements, we can target a less massive star. As stated in \S \ref{S:Intro}, this method will be optimal for stars for which evolutionary models are less able to constrain stellar parameters precisely, such as F-type stars, subgiants, and M-dwarfs. Since M-dwarfs are less massive than their G-type counterparts, a given mass planet around an M-dwarf produces a comparatively larger RV signal. Since the mass uncertainties will generally be dominated by the Doppler uncertainty, focusing on low-mass stars will enhance the observed signal, allowing for more meaningful mass constraints to be set. As proof, we again simulate observations of orbiting planets, but with larger values for $K$, corresponding to a less massive star or more massive planets. By sampling at the same times and assuming the same statistical errors and jitter levels, the fractional error in $K$ decreases significantly for a fixed number of observations. These results are also shown in Fig. \ref{RVerrors} (dashed lines). For example, a planet identical to Kepler-18 c orbiting a star of mass $M = 0.33 M_\odot$ would produce a semiamplitude $K = 15$ m s$^{-1}$; with only 20 observations the RV semiamplitude could be constrained to within 8 percent and the stellar mass to within 30 percent. It is worth noting that these observations are all simulated assuming similar levels of statistical noise as the Kepler-18 observations. This is a reasonable approximation for the stars hosting Kepler Objects of Interest, but these stars are considerably fainter than the average Doppler planet search target. If transit timing variations are detected around a considerably brighter star, as one would expect from next-generation space-based planet finding missions, radial velocity observations could be carried out to considerably higher precision, decreasing the number of observations required to precisely measure the stellar radial velocity semiamplitude.

\section{Summary and Discussion}

\label{S:SD}

We present a  method of measuring stellar and planetary masses dynamically by combining TTVs measured from transit light curves and follow-up radial velocity measurements. Our method can be used as an alternative to relying on stellar evolutionary models, which can be poorly constrained for non-solar type stars such as M-dwarfs, subgiants, and F stars. By analyzing the Kepler-18 system and confirming our expressions with dynamical simulations of this system, we show the potential of our method.

While we show our method to be viable, especially for low-mass stars, using our method requires a somewhat specific set of circumstances. The system must contain two planets with masses large enough to force a detectable Doppler signal and observable transit timing variations on circular orbits near a first-order commensurability. \Kepler data suggests planets near resonance are common: more than 12 percent of planet systems show evidence for detectable transit timing variations \citep{Ford12a}, and dozens of planets near resonance have been confirmed through TTVs \citep{Steffen12b}. Both the TTV and Doppler signals can be measured for super-Earth planets with periods less than 30 days; short-period systems such as these are extremely common \citep{Howard12}. Thus, it is likely that despite the specific requirements needed to use our system, it can be applied to a considerable number of \Kepler planetary systems.

As shown in Equations 14 and 15 of L12, the amplitude of the TTV signal is given such that
\begin{equation}
|V| \sim |V_\textrm{damped}|\bigg(1 + \frac{|Z_\textrm{free}|}{|\Delta|}\bigg),
\end{equation}
with $|V_\textrm{damped}|$ the amplitude of the TTV signal if the system were damped of its free eccentricity. The quantity $Z_\textrm{free}$ is defined such that $Z_\textrm{free} = f z_\textrm{free} + g z^\prime_\textrm{free}$, where $z$ is the complex eccentricity of the planet, $z = e\exp(i \omega)$. Thus, our method as described will break down unless $|Z_{\textrm{free}}| \ll |\Delta|$. This is a reasonable assumption for planets with periods under ten days. In theory, even if a non-negligible amount of free eccentricity remains in the system, a detailed radial velocity orbital solution and light curve analysis could be used to calculate $Z_{\textrm{free}}$ and determine the stellar and planetary masses.

As a projection of the utility of this method, consider KOI 1241, a system containing two planets with periods of 10.5 and 21.4 days orbiting a giant star \citep[$R = 3.14 R_\odot$,][]{Steffen12b}. From the transit timing signal, there is evidence that this system has not dissipated its free eccentricity, meaning it is not optimal for our study. However, with nine quarters of public \Kepler data, we can constrain the TTV signal caused by the larger planet in this system to 8.2 percent. Moreover, with only nine radial velocity observations, the radial velocity semiamplitude of the larger planet can be measured to 5.3 percent. Thus, from our method alone, if a system existed that was nearly identical to KOI 1241 but damped of free eccentricity, by Equations 11-13, we expect we could determine the stellar mass to 29 percent and the mass of the larger planet to within 22 percent. Our method could also be applied to KOI 1241 in the future if enough radial velocity data is collected to determine the magnitude of $Z_\textrm{free}$ for the system. The uncertainties in the TTV signal of KOI 1241 are larger than the uncertainty in the RV semiamplitude. The transit timing errors will decrease as more \Kepler data are released: decreasing the TTV error to five percent without including any additional radial velocity observations will reduce the uncertainty in the stellar mass to 20 percent. Thus our method could provide significant constraints on stellar masses in regimes where stellar atmospheres are less well-understood, such as subgiants and cool stars. For these cases, our method will be able to compliment asteroseismology results as an independent measure on the mass of the star. Moreover, our method can be used to find systems where the analytic stellar mass is substantially different than the Kepler Input Catalog values, which can then be followed up with dynamical modeling, asteroseismology, or high resolution spectroscopy to better characterize the star and orbiting planets.

Our method of estimating the stellar mass of a planet host star requires a many high-precision radial velocity measurements as well as observations of enough measurements to measure TTVs. Traditionally, for radial velocity searches the stellar mass is estimated by interpolating spectroscopic properties such as the stellar effective temperature and metallicity onto stellar evolution model grids \citep[e.g.][]{Sozzetti07}. From a careful analysis of transit light curves, one can measure the ratio $a/R_\ast$ and the duration of the transit event, which provide an estimate of the stellar density. However, stellar evolution models are still required to break the degeneracy between stellar radius and mass. While these traditional methods allow for precise measurements of stellar masses, the results are model-dependent and may be subject to systematic errors induced by the stellar models. This is particularly true for stars very different from the Sun. Our method has the advantage of providing an accurate, model-independent stellar mass, albeit with a trade-off in precision. However, even in cases for which the precision of our method is poor, the resulting mass estimate can be used as a prior in the more traditional model-grid interpolation.  

With present data our technique is only viable as an alternative to stellar modeling in the most exceptional cases. Transit timing variations have been detected to remarkable precision by \textit{Kepler}, but very few KOIs have been followed up with radial velocity measurements. In the cases where RV data exists, only enough measurements were collected to confirm the planetary nature of the system, not to independently measure the planetary masses \citep{Holman10}. Our routine will become more useful for systems around brighter stars, when more RV measurements can be efficiently collected and the RV semiamplitude can be better constrained. 

Despite the faintness of the \Kepler planet candidate host stars, there are a few stars that would be ideal candidates for applications of our method. From the collection of Kepler Objects of Interest, we searched for stars hosting at least two transiting planets each with $P < 25$ days. We required at least one planet to be larger than 2 $R_\oplus$ and the planet periods to lie within five percent of a first-order mean-motion resonance. To ensure that all targets were optimized for radial velocity follow-up, we eliminated all targets fainter than $m_{\textrm{Kp}} = 13.0$. After making these cuts, we find 8 candidate systems to which this technique can be applied: KOI 85, 111, 115, 117, 244, 304, 1241 and 1930. As stated earlier in this section, KOI 1241 is not an ideal target because it has not been fully damped of its primordial eccentricity and there is not enough radial velocity information to uniquely determine the eccentricity of both planets. Of the remaining 7 systems, the CPS team has collected more than 10 radial velocity measurements only on one, KOI 244. Additional radial velocity measurements of any or all of the above systems would enable further validation of our procedure as well as additional constraints on the masses of each of the stars and their planets. Moreover, next-generation planet finding missions, such as TESS \citep{Brown08} and PLATO \citep{Catala10} will target bright stars, making detailed radial velocity follow-up observations of systems exhibiting transit timing variations a much more practical possibility. 

\acknowledgements{We thank Rebekah Dawson, Jonathan Swift, Daniel Fabrycky, and Michael Bottom for insightful discussions and helpful comments during the preparation of this manuscript. We also thank Philip Muirhead for thoughtful commentary and discussion during the July 2012 Sagan Exoplanet Summer Workshop. The data presented herein were obtained in part at the W. M. Keck Observatory, which is operated as a scientific partnership among the California Institute of Technology, the University of California, and NASA, and was made possible by the generous financial support of the W. M. Keck Foundation. We extend special thanks to those of Hawaiian ancestry on whose sacred mountain of Mauna Kea we are privileged to be guests. Without their generous hospitality, the HIRES observations presented herein would not have been possible.

\LongTables
\begin{deluxetable}{lcccc}
\tablewidth{0pt}
\tabletypesize{\footnotesize}
\tablecaption{Transit Times for Kepler Transiting Planet Candidates\label{tabTTs}}
\tablehead{
\colhead{KOI} & \colhead{ n } & \colhead{ $t_n$}    &  \colhead{TTV$_n$} & \colhead{ $\sigma_n$} \\ 
   \colhead{}  &   \colhead{}     &  \colhead{BJD-2454900} &  \colhead{(d)}  &  \colhead{(d)} }
\startdata

137.01 & 0 & 198.3142 & -0.0006 &  0.0012 \\
137.01 & 1 & 205.9557 &  0.0002 &  0.0013 \\
137.01 & 2 & 213.5973 &  0.0019 &  0.0018 \\
137.01 & 3 & 221.2389 &  0.0017 &  0.0010 \\
137.01 & 4 & 228.8804 &  0.0021 &  0.0017 \\
137.01 & 5 & 236.5220 &  0.0014 &  0.0015 \\
137.01 & 6 & 244.1636 &  0.0031 &  0.0011 \\
137.01 & 7 & 251.8052 &  0.0030 &  0.0011 \\
137.01 & 8 & 259.4467 &  0.0037 &  0.0012 \\
137.01 & 9 & 267.0883 &  0.0050 &  0.0012 \\
137.01 & 10 & 274.7299 &  0.0041 &  0.0012 \\
137.01 & 11 & 282.3714 &  0.0037 &  0.0018 \\
137.01 & 12 & 290.0130 &  0.0035 &  0.0018 \\
137.01 & 13 & 297.6546 &  0.0028 &  0.0012 \\
137.01 & 14 & 305.2961 &  0.0034 &  0.0011 \\
137.01 & 15 & 312.9377 &  0.0026 &  0.0011 \\
137.01 & 16 & 320.5793 &  0.0013 &  0.0011 \\
137.01 & 17 & 328.2209 &  0.0018 &  0.0012 \\
137.01 & 18 & 335.8624 &  0.0008 &  0.0015 \\
137.01 & 20 & 351.1456 &  0.0001 &  0.0011 \\
137.01 & 21 & 358.7871 & -0.0005 &  0.0012 \\
137.01 & 22 & 366.4287 & -0.0027 &  0.0014 \\
137.01 & 23 & 374.0703 & -0.0033 &  0.0014 \\
137.01 & 24 & 381.7119 & -0.0024 &  0.0011 \\
137.01 & 25 & 389.3534 & -0.0025 &  0.0011 \\
137.01 & 26 & 396.9950 & -0.0033 &  0.0011 \\
137.01 & 27 & 404.6366 & -0.0036 &  0.0013 \\
137.01 & 28 & 412.2781 & -0.0046 &  0.0012 \\
137.01 & 29 & 419.9197 & -0.0037 &  0.0011 \\
137.01 & 30 & 427.5613 & -0.0038 &  0.0015 \\
137.01 & 31 & 435.2028 & -0.0039 &  0.0013 \\
137.01 & 32 & 442.8444 & -0.0019 &  0.0012 \\
137.01 & 33 & 450.4860 & -0.0013 &  0.0011 \\
137.01 & 34 & 458.1276 & -0.0017 &  0.0010 \\
137.01 & 35 & 465.7691 & -0.0008 &  0.0015 \\
137.01 & 36 & 473.4107 &  0.0005 &  0.0013 \\
137.01 & 37 & 481.0523 &  0.0016 &  0.0011 \\
137.01 & 38 & 488.6938 &  0.0026 &  0.0014 \\
137.01 & 39 & 496.3354 &  0.0027 &  0.0015 \\
137.01 & 40 & 503.9770 &  0.0030 &  0.0014 \\
137.01 & 41 & 511.6186 &  0.0038 &  0.0011 \\
137.01 & 42 & 519.2601 &  0.0032 &  0.0013 \\
137.01 & 43 & 526.9017 &  0.0052 &  0.0013 \\
137.01 & 44 & 534.5433 &  0.0055 &  0.0013 \\
137.01 & 45 & 542.1848 &  0.0034 &  0.0010 \\
137.01 & 46 & 549.8264 &  0.0025 &  0.0014 \\
137.01 & 47 & 557.4680 &  0.0049 &  0.0021 \\
137.01 & 48 & 565.1095 &  0.0035 &  0.0011 \\
137.01 & 49 & 572.7511 &  0.0030 &  0.0012 \\
137.01 & 50 & 580.3927 &  0.0017 &  0.0010 \\
137.01 & 51 & 588.0343 &  0.0034 &  0.0016 \\
137.01 & 52 & 595.6758 &  0.0013 &  0.0011 \\
137.01 & 54 & 610.9590 & -0.0006 &  0.0012 \\
137.01 & 55 & 618.6005 & -0.0019 &  0.0012 \\
137.01 & 56 & 626.2421 & -0.0015 &  0.0016 \\
137.01 & 57 & 633.8837 & -0.0026 &  0.0012 \\
137.01 & 58 & 641.5253 & -0.0016 &  0.0012 \\
137.01 & 61 & 664.4500 & -0.0046 &  0.0014 \\
137.01 & 62 & 672.0915 & -0.0028 &  0.0012 \\
137.01 & 63 & 679.7331 & -0.0042 &  0.0012 \\
137.01 & 65 & 695.0162 & -0.0032 &  0.0011 \\
137.01 & 66 & 702.6578 & -0.0025 &  0.0013 \\
137.01 & 67 & 710.2994 & -0.0029 &  0.0012 \\
137.01 & 68 & 717.9410 & -0.0029 &  0.0011 \\
137.01 & 69 & 725.5825 & -0.0004 &  0.0013 \\
137.01 & 71 & 740.8657 &  0.0006 &  0.0013 \\
137.01 & 72 & 748.5072 &  0.0006 &  0.0014 \\
137.01 & 73 & 756.1488 &  0.0014 &  0.0011 \\
137.01 & 74 & 763.7904 &  0.0049 &  0.0017 \\
137.01 & 75 & 771.4320 &  0.0011 &  0.0016 \\
137.01 & 76 & 779.0735 &  0.0033 &  0.0012 \\
137.01 & 77 & 786.7151 &  0.0021 &  0.0011 \\
137.01 & 78 & 794.3567 &  0.0031 &  0.0012 \\
137.01 & 79 & 801.9982 &  0.0039 &  0.0012 \\
137.01 & 80 & 809.6398 &  0.0029 &  0.0014 \\
137.01 & 81 & 817.2814 &  0.0038 &  0.0011 \\
137.01 & 82 & 824.9229 &  0.0016 &  0.0012 \\
137.02 & 0 & 194.8832 &  0.0000 &  0.0010 \\
137.02 & 1 & 209.7421 & -0.0012 &  0.0011 \\
137.02 & 2 & 224.6011 & -0.0025 &  0.0010 \\
137.02 & 3 & 239.4600 & -0.0037 &  0.0010 \\
137.02 & 4 & 254.3189 & -0.0029 &  0.0010 \\
137.02 & 6 & 284.0368 & -0.0032 &  0.0011 \\
137.02 & 7 & 298.8958 & -0.0040 &  0.0011 \\
137.02 & 8 & 313.7547 & -0.0013 &  0.0010 \\
137.02 & 9 & 328.6136 & -0.0023 &  0.0023 \\
137.02 & 10 & 343.4726 & -0.0005 &  0.0011 \\
137.02 & 11 & 358.3315 &  0.0008 &  0.0009 \\
137.02 & 12 & 373.1905 &  0.0017 &  0.0012 \\
137.02 & 13 & 388.0494 &  0.0036 &  0.0010 \\
137.02 & 14 & 402.9083 &  0.0029 &  0.0011 \\
137.02 & 15 & 417.7673 &  0.0020 &  0.0010 \\
137.02 & 16 & 432.6262 &  0.0008 &  0.0012 \\
137.02 & 17 & 447.4852 &  0.0008 &  0.0011 \\
137.02 & 18 & 462.3441 &  0.0008 &  0.0011 \\
137.02 & 19 & 477.2030 & -0.0001 &  0.0011 \\
137.02 & 20 & 492.0620 &  0.0002 &  0.0010 \\
137.02 & 21 & 506.9209 & -0.0023 &  0.0011 \\
137.02 & 22 & 521.7799 & -0.0024 &  0.0010 \\
137.02 & 23 & 536.6388 & -0.0038 &  0.0010 \\
137.02 & 24 & 551.4977 & -0.0035 &  0.0010 \\
137.02 & 25 & 566.3567 & -0.0033 &  0.0011 \\
137.02 & 26 & 581.2156 & -0.0033 &  0.0010 \\
137.02 & 27 & 596.0746 & -0.0013 &  0.0015 \\
137.02 & 29 & 625.7924 &  0.0020 &  0.0010 \\
137.02 & 31 & 655.5103 &  0.0032 &  0.0014 \\
137.02 & 32 & 670.3693 &  0.0015 &  0.0011 \\
137.02 & 33 & 685.2282 &  0.0026 &  0.0009 \\
137.02 & 34 & 700.0871 & -0.0011 &  0.0011 \\
137.02 & 35 & 714.9461 &  0.0008 &  0.0010 \\
137.02 & 36 & 729.8050 & -0.0019 &  0.0010 \\
137.02 & 37 & 744.6640 & -0.0012 &  0.0011 \\
137.02 & 38 & 759.5229 & -0.0027 &  0.0011 \\
137.02 & 39 & 774.3818 & -0.0030 &  0.0012 \\
137.02 & 40 & 789.2408 & -0.0034 &  0.0010 \\
137.02 & 41 & 804.0997 & -0.0052 &  0.0010 \\
137.02 & 42 & 818.9587 &  0.0003 &  0.0013 \\

\enddata
\label{TTTable}
\end{deluxetable}


\begin{thebibliography}{33}
\expandafter\ifx\csname natexlab\endcsname\relax\def\natexlab#1{#1}\fi

\bibitem[{{Agol} {et~al.}(2005){Agol}, {Steffen}, {Sari}, \&
  {Clarkson}}]{Agol05}
{Agol}, E., {Steffen}, J., {Sari}, R., \& {Clarkson}, W. 2005, \mnras, 359, 567

\bibitem[{{Ballard} {et~al.}(2011){Ballard}, {Fabrycky}, {Fressin},
  {Charbonneau}, {Desert}, {Torres}, {Marcy}, {Burke}, {Isaacson}, {Henze},
  {Steffen}, {Ciardi}, {Howell}, {Cochran}, {Endl}, {Bryson}, {Rowe}, {Holman},
  {Lissauer}, {Jenkins}, {Still}, {Ford}, {Christiansen}, {Middour}, {Haas},
  {Li}, {Hall}, {McCauliff}, {Batalha}, {Koch}, \& {Borucki}}]{Ballard11}
{Ballard}, S., {Fabrycky}, D., {Fressin}, F., {et~al.} 2011, \apj, 743, 200

\bibitem[{{Batalha} {et~al.}(2012){Batalha}, {Rowe}, {Bryson}, {Barclay},
  {Burke}, {Caldwell}, {Christiansen}, {Mullally}, {Thompson}, {Brown},
  {Dupree}, {Fabrycky}, {Ford}, {Fortney}, {Gilliland}, {Isaacson}, {Latham},
  {Marcy}, {Quinn}, {Ragozzine}, {Shporer}, {Borucki}, {Ciardi}, {Gautier},
  {Haas}, {Jenkins}, {Koch}, {Lissauer}, {Rapin}, {Basri}, {Boss}, {Buchhave},
  {Charbonneau}, {Christensen-Dalsgaard}, {Clarke}, {Cochran}, {Demory},
  {Devore}, {Esquerdo}, {Everett}, {Fressin}, {Geary}, {Girouard}, {Gould},
  {Hall}, {Holman}, {Howard}, {Howell}, {Ibrahim}, {Kinemuchi}, {Kjeldsen},
  {Klaus}, {Li}, {Lucas}, {Morris}, {Prsa}, {Quintana}, {Sanderfer},
  {Sasselov}, {Seader}, {Smith}, {Steffen}, {Still}, {Stumpe}, {Tarter},
  {Tenenbaum}, {Torres}, {Twicken}, {Uddin}, {Van Cleve}, {Walkowicz}, \&
  {Welsh}}]{Batalha12}
{Batalha}, N.~M., {Rowe}, J.~F., {Bryson}, S.~T., {et~al.} 2012, arXiv:1202.5852

\bibitem[{{Borucki} {et~al.}(2011){Borucki}, {Koch}, {Basri}, {Batalha},
  {Brown}, {Bryson}, {Caldwell}, {Christensen-Dalsgaard}, {Cochran}, {DeVore},
  {Dunham}, {Gautier}, {Geary}, {Gilliland}, {Gould}, {Howell}, {Jenkins},
  {Latham}, {Lissauer}, {Marcy}, {Rowe}, {Sasselov}, {Boss}, {Charbonneau},
  {Ciardi}, {Doyle}, {Dupree}, {Ford}, {Fortney}, {Holman}, {Seager},
  {Steffen}, {Tarter}, {Welsh}, {Allen}, {Buchhave}, {Christiansen}, {Clarke},
  {Das}, {D{\'e}sert}, {Endl}, {Fabrycky}, {Fressin}, {Haas}, {Horch},
  {Howard}, {Isaacson}, {Kjeldsen}, {Kolodziejczak}, {Kulesa}, {Li}, {Lucas},
  {Machalek}, {McCarthy}, {MacQueen}, {Meibom}, {Miquel}, {Prsa}, {Quinn},
  {Quintana}, {Ragozzine}, {Sherry}, {Shporer}, {Tenenbaum}, {Torres},
  {Twicken}, {Van Cleve}, {Walkowicz}, {Witteborn}, \& {Still}}]{Borucki11}
{Borucki}, W.~J., {Koch}, D.~G., {Basri}, G., {et~al.} 2011, \apj, 736, 19

\bibitem[{{Brown} \& {Latham}(2008)}]{Brown08}
{Brown}, T.~M., \& {Latham}, D.~W. 2008, arXiv:0812.1305

\bibitem[{{Catala} {et~al.}(2010){Catala}, {Arentoft}, {Fridlund}, {Lindberg},
  {Mas-Hesse}, {Micela}, {Pollacco}, {Poretti}, {Rauer}, {Roxburgh}, {Stankov},
  \& {Udry}}]{Catala10}
{Catala}, C., {Arentoft}, T., {Fridlund}, M., {et~al.} 2010, in Astronomical
  Society of the Pacific Conference Series, Vol. 430, Pathways Towards
  Habitable Planets, ed. V.~{Coud{\'e} Du Foresto}, D.~M. {Gelino}, \&
  I.~{Ribas}, 260

\bibitem[{{Cochran} {et~al.}(2011){Cochran}, {Fabrycky}, {Torres}, {Fressin},
  {D{\'e}sert}, {Ragozzine}, {Sasselov}, {Fortney}, {Rowe}, {Brugamyer},
  {Bryson}, {Carter}, {Ciardi}, {Howell}, {Steffen}, {Borucki}, {Koch}, {Winn},
  {Welsh}, {Uddin}, {Tenenbaum}, {Still}, {Seager}, {Quinn}, {Mullally},
  {Miller}, {Marcy}, {MacQueen}, {Lucas}, {Lissauer}, {Latham}, {Knutson},
  {Kinemuchi}, {Johnson}, {Jenkins}, {Isaacson}, {Howard}, {Horch}, {Holman},
  {Henze}, {Haas}, {Gilliland}, {Gautier}, {Ford}, {Fischer}, {Everett},
  {Endl}, {Demory}, {Deming}, {Charbonneau}, {Caldwell}, {Buchhave}, {Brown},
  \& {Batalha}}]{Cochran11}
{Cochran}, W.~D., {Fabrycky}, D.~C., {Torres}, G., {et~al.} 2011, \apjs, 197, 7

\bibitem[{{Dawson} \& {Johnson}(2012)}]{DawsonJJ12}
{Dawson}, R.~I., \& {Johnson}, J.~A. 2012, \apj, 756, 122

\bibitem[{{Dawson} {et~al.}(2012){Dawson}, {Johnson}, {Morton}, {Crepp},
  {Fabrycky}, {Murray-Clay}, \& {Howard}}]{Dawson12}
{Dawson}, R.~I., {Johnson}, J.~A., {Morton}, T.~D., {et~al.} 2012, arXiv:1206.5579

\bibitem[{{Doyle} {et~al.}(2011){Doyle}, {Carter}, {Fabrycky}, {Slawson},
  {Howell}, {Winn}, {Orosz}, {Pr{\'}sa}, {Welsh}, {Quinn}, {Latham}, {Torres},
  {Buchhave}, {Marcy}, {Fortney}, {Shporer}, {Ford}, {Lissauer}, {Ragozzine},
  {Rucker}, {Batalha}, {Jenkins}, {Borucki}, {Koch}, {Middour}, {Hall},
  {McCauliff}, {Fanelli}, {Quintana}, {Holman}, {Caldwell}, {Still},
  {Stefanik}, {Brown}, {Esquerdo}, {Tang}, {Furesz}, {Geary}, {Berlind},
  {Calkins}, {Short}, {Steffen}, {Sasselov}, {Dunham}, {Cochran}, {Boss},
  {Haas}, {Buzasi}, \& {Fischer}}]{Doyle11}
{Doyle}, L.~R., {Carter}, J.~A., {Fabrycky}, D.~C., {et~al.} 2011, Science,
  333, 1602

\bibitem[{{Fabrycky} {et~al.}(2012){Fabrycky}, {Ford}, {Steffen}, {Rowe},
  {Carter}, {Moorhead}, {Batalha}, {Borucki}, {Bryson}, {Buchhave},
  {Christiansen}, {Ciardi}, {Cochran}, {Endl}, {Fanelli}, {Fischer}, {Fressin},
  {Geary}, {Haas}, {Hall}, {Holman}, {Jenkins}, {Koch}, {Latham}, {Li},
  {Lissauer}, {Lucas}, {Marcy}, {Mazeh}, {McCauliff}, {Quinn}, {Ragozzine},
  {Sasselov}, \& {Shporer}}]{Fabrycky12}
{Fabrycky}, D.~C., {Ford}, E.~B., {Steffen}, J.~H., {et~al.} 2012, \apj, 750,
  114

\bibitem[{{Ford} {et~al.}(2012{\natexlab{a}}){Ford}, {Fabrycky}, {Steffen},
  {Carter}, {Fressin}, {Holman}, {Lissauer}, {Moorhead}, {Morehead},
  {Ragozzine}, {Rowe}, {Welsh}, {Allen}, {Batalha}, {Borucki}, {Bryson},
  {Buchhave}, {Burke}, {Caldwell}, {Charbonneau}, {Clarke}, {Cochran},
  {D{\'e}sert}, {Endl}, {Everett}, {Fischer}, {Gautier}, {Gilliland},
  {Jenkins}, {Haas}, {Horch}, {Howell}, {Ibrahim}, {Isaacson}, {Koch},
  {Latham}, {Li}, {Lucas}, {MacQueen}, {Marcy}, {McCauliff}, {Mullally},
  {Quinn}, {Quintana}, {Shporer}, {Still}, {Tenenbaum}, {Thompson}, {Torres},
  {Twicken}, {Wohler}, \& {the Kepler Science Team}}]{Ford12TTV}
{Ford}, E.~B., {Fabrycky}, D.~C., {Steffen}, J.~H., {et~al.}
  2012{\natexlab{a}}, \apj, 750, 113

\bibitem[{{Ford} {et~al.}(2012{\natexlab{b}}){Ford}, {Ragozzine}, {Rowe},
  {Steffen}, {Barclay}, {Batalha}, {Borucki}, {Bryson}, {Caldwell}, {Fabrycky},
  {Gautier}, {Holman}, {Ibrahim}, {Kjeldsen}, {Kinemuchi}, {Koch}, {Lissauer},
  {Still}, {Tenenbaum}, {Uddin}, \& {Welsh}}]{Ford12a}
{Ford}, E.~B., {Ragozzine}, D., {Rowe}, J.~F., {et~al.} 2012{\natexlab{b}},
  arXiv:1201.1892

\bibitem[{{Holman} {et~al.}(2010){Holman}, {Fabrycky}, {Ragozzine}, {Ford},
  {Steffen}, {Welsh}, {Lissauer}, {Latham}, {Marcy}, {Walkowicz}, {Batalha},
  {Jenkins}, {Rowe}, {Cochran}, {Fressin}, {Torres}, {Buchhave}, {Sasselov},
  {Borucki}, {Koch}, {Basri}, {Brown}, {Caldwell}, {Charbonneau}, {Dunham},
  {Gautier}, {Geary}, {Gilliland}, {Haas}, {Howell}, {Ciardi}, {Endl},
  {Fischer}, {F{\"u}r{\'e}sz}, {Hartman}, {Isaacson}, {Johnson}, {MacQueen},
  {Moorhead}, {Morehead}, \& {Orosz}}]{Holman10}
{Holman}, M.~J., {Fabrycky}, D.~C., {Ragozzine}, D., {et~al.} 2010, Science,
  330, 51

\bibitem[{{Howard} {et~al.}(2012){Howard}, {Marcy}, {Bryson}, {Jenkins},
  {Rowe}, {Batalha}, {Borucki}, {Koch}, {Dunham}, {Gautier}, {Van Cleve},
  {Cochran}, {Latham}, {Lissauer}, {Torres}, {Brown}, {Gilliland}, {Buchhave},
  {Caldwell}, {Christensen-Dalsgaard}, {Ciardi}, {Fressin}, {Haas}, {Howell},
  {Kjeldsen}, {Seager}, {Rogers}, {Sasselov}, {Steffen}, {Basri},
  {Charbonneau}, {Christiansen}, {Clarke}, {Dupree}, {Fabrycky}, {Fischer},
  {Ford}, {Fortney}, {Tarter}, {Girouard}, {Holman}, {Johnson}, {Klaus},
  {Machalek}, {Moorhead}, {Morehead}, {Ragozzine}, {Tenenbaum}, {Twicken},
  {Quinn}, {Isaacson}, {Shporer}, {Lucas}, {Walkowicz}, {Welsh}, {Boss},
  {Devore}, {Gould}, {Smith}, {Morris}, {Prsa}, {Morton}, {Still}, {Thompson},
  {Mullally}, {Endl}, \& {MacQueen}}]{Howard12}
{Howard}, A.~W., {Marcy}, G.~W., {Bryson}, S.~T., {et~al.} 2012, \apjs, 201, 15

\bibitem[{{Jha} {et~al.}(2000){Jha}, {Charbonneau}, {Garnavich}, {Sullivan},
  {Sullivan}, {Brown}, \& {Tonry}}]{Jha00}
{Jha}, S., {Charbonneau}, D., {Garnavich}, P.~M., {et~al.} 2000, \apjl, 540,
  L45

\bibitem[{{Johnson} {et~al.}(2012{\natexlab{a}}){Johnson}, {Morton}, \&
  {Wright}}]{JohnsonMW12}
{Johnson}, J.~A., {Morton}, T.~D., \& {Wright}, J.~T. 2012{\natexlab{a}}, arXiv:1208.4377

\bibitem[{{Johnson} {et~al.}(2012{\natexlab{b}}){Johnson}, {Gazak}, {Apps},
  {Muirhead}, {Crepp}, {Crossfield}, {Boyajian}, {von Braun}, {Rojas-Ayala},
  {Howard}, {Covey}, {Schlawin}, {Hamren}, {Morton}, {Marcy}, \&
  {Lloyd}}]{KOI254}
{Johnson}, J.~A., {Gazak}, J.~Z., {Apps}, K., {et~al.} 2012{\natexlab{b}}, \aj,
  143, 111

\bibitem[{{Kipping}(2010)}]{Kipping10}
{Kipping}, D.~M. 2010, \mnras, 409, L119

\bibitem[{{Lissauer} {et~al.}(2011){Lissauer}, {Fabrycky}, {Ford}, {Borucki},
  {Fressin}, {Marcy}, {Orosz}, {Rowe}, {Torres}, {Welsh}, {Batalha}, {Bryson},
  {Buchhave}, {Caldwell}, {Carter}, {Charbonneau}, {Christiansen}, {Cochran},
  {Desert}, {Dunham}, {Fanelli}, {Fortney}, {Gautier}, {Geary}, {Gilliland},
  {Haas}, {Hall}, {Holman}, {Koch}, {Latham}, {Lopez}, {McCauliff}, {Miller},
  {Morehead}, {Quintana}, {Ragozzine}, {Sasselov}, {Short}, \&
  {Steffen}}]{Lissauer11}
{Lissauer}, J.~J., {Fabrycky}, D.~C., {Ford}, E.~B., {et~al.} 2011, \nat, 470,
  53

\bibitem[{{Lissauer} {et~al.}(2012){Lissauer}, {Marcy}, {Rowe}, {Bryson},
  {Adams}, {Buchhave}, {Ciardi}, {Cochran}, {Fabrycky}, {Ford}, {Fressin},
  {Geary}, {Gilliland}, {Holman}, {Howell}, {Jenkins}, {Kinemuchi}, {Koch},
  {Morehead}, {Ragozzine}, {Seader}, {Tanenbaum}, {Torres}, \&
  {Twicken}}]{Lissauer12}
{Lissauer}, J.~J., {Marcy}, G.~W., {Rowe}, J.~F., {et~al.} 2012, \apj, 750, 112

\bibitem[{{Lithwick} {et~al.}(2012){Lithwick}, {Xie}, \& {Wu}}]{Lithwick12}
{Lithwick}, Y., {Xie}, J., \& {Wu}, Y. 2012, arXiv:1207.4192

\bibitem[{{Mandel} \& {Agol}(2002)}]{Mandel02}
{Mandel}, K., \& {Agol}, E. 2002, \apjl, 580, L171

\bibitem[{{Meschiari} {et~al.}(2009){Meschiari}, {Wolf}, {Rivera}, {Laughlin},
  {Vogt}, \& {Butler}}]{Meschiari09}
{Meschiari}, S., {Wolf}, A.~S., {Rivera}, E., {et~al.} 2009, \pasp, 121, 1016

\bibitem[{{Muirhead} {et~al.}(2012){Muirhead}, {Johnson}, {Apps}, {Carter},
  {Morton}, {Fabrycky}, {Pineda}, {Bottom}, {Rojas-Ayala}, {Schlawin},
  {Hamren}, {Covey}, {Crepp}, {Stassun}, {Pepper}, {Hebb}, {Kirby}, {Howard},
  {Isaacson}, {Marcy}, {Levitan}, {Diaz-Santos}, {Armus}, \& {Lloyd}}]{KOI961}
{Muirhead}, P.~S., {Johnson}, J.~A., {Apps}, K., {et~al.} 2012, \apj, 747, 144

\bibitem[{{Murray} \& {Dermott}(1999)}]{MDSSD}
{Murray}, C.~D., \& {Dermott}, S.~F. 1999, {Solar System Dynamics} (Cambridge
  University Press)

\bibitem[{{Nesvorn{\'y}} {et~al.}(2012){Nesvorn{\'y}}, {Kipping}, {Buchhave},
  {Bakos}, {Hartman}, \& {Schmitt}}]{Nesvorny12}
{Nesvorn{\'y}}, D., {Kipping}, D.~M., {Buchhave}, L.~A., {et~al.} 2012,
  Science, 336, 1133

\bibitem[{{Paddock}(1913)}]{Paddock13}
{Paddock}, G.~F. 1913, \pasp, 25, 208

\bibitem[{{Sozzetti} {et~al.}(2007){Sozzetti}, {Torres}, {Charbonneau},
  {Latham}, {Holman}, {Winn}, {Laird}, \& {O'Donovan}}]{Sozzetti07}
{Sozzetti}, A., {Torres}, G., {Charbonneau}, D., {et~al.} 2007, \apj, 664, 1190

\bibitem[{{Steffen} {et~al.}(2012{\natexlab{a}}){Steffen}, {Ford}, {Rowe},
  {Fabrycky}, {Holman}, {Welsh}, {Borucki}, {Batalha}, {Bryson}, {Caldwell},
  {Ciardi}, {Jenkins}, {Kjeldsen}, {Koch}, {Prsa}, {Sanderfer}, {Seader}, \&
  {Twicken}}]{Steffen12a}
{Steffen}, J.~H., {Ford}, E.~B., {Rowe}, J.~F., {et~al.} 2012{\natexlab{a}},
  arXiv:1201.1873

\bibitem[{{Steffen} {et~al.}(2012{\natexlab{b}}){Steffen}, {Fabrycky}, {Agol},
  {Ford}, {Morehead}, {Cochran}, {Lissauer}, {Adams}, {Borucki}, {Bryson},
  {Caldwell}, {Dupree}, {Jenkins}, {Robertson}, {Rowe}, {Seader}, {Thompson},
  \& {Twicken}}]{Steffen12b}
{Steffen}, J.~H., {Fabrycky}, D.~C., {Agol}, E., {et~al.} 2012{\natexlab{b}},
  arXiv:1208.3499

\bibitem[{{Valenti} \& {Fischer}(2005)}]{ValFis05}
{Valenti}, J.~A., \& {Fischer}, D.~A. 2005, \apjs, 159, 141

\bibitem[{{Welsh} {et~al.}(2012){Welsh}, {Orosz}, {Carter}, {Fabrycky}, {Ford},
  {Lissauer}, {Pr{\v s}a}, {Quinn}, {Ragozzine}, {Short}, {Torres}, {Winn},
  {Doyle}, {Barclay}, {Batalha}, {Bloemen}, {Brugamyer}, {Buchhave},
  {Caldwell}, {Caldwell}, {Christiansen}, {Ciardi}, {Cochran}, {Endl},
  {Fortney}, {Gautier}, {Gilliland}, {Haas}, {Hall}, {Holman}, {Howard},
  {Howell}, {Isaacson}, {Jenkins}, {Klaus}, {Latham}, {Li}, {Marcy}, {Mazeh},
  {Quintana}, {Robertson}, {Shporer}, {Steffen}, {Windmiller}, {Koch}, \&
  {Borucki}}]{Welsh12}
{Welsh}, W.~F., {Orosz}, J.~A., {Carter}, J.~A., {et~al.} 2012, \nat, 481, 475

\end{thebibliography}
\end{document}